\documentclass[submission,copyright,creativecommons]{eptcs}
\providecommand{\event}{HAS 2013} 
\usepackage{breakurl}             

\usepackage{color}
\usepackage{array}
\usepackage{wrapfig}
\usepackage{amsmath,amssymb}
\usepackage{graphicx}
\usepackage{faktor}
\usepackage[caption=false,font=footnotesize]{subfig}

\newtheorem{theorem}{Theorem}[section]
\newtheorem{defn}[theorem]{Definition}
\newtheorem{prop}[theorem]{Property}

\newtheorem{proposition}[theorem]{Proposition}
\newtheorem{example}[theorem]{Example}

\title{Sampling-based Decentralized Monitoring\\ for Networked Embedded Systems}
\author{Ezio Bartocci
\institute{Institute of Computer Engineering\\
Vienna University of Technology\\
Vienna, Austria}
\email{ezio.bartocci@tuwien.ac.at}
}
\def\titlerunning{Sampling-based Decentralized Monitoring for Networked Embedded Systems}

\def\authorrunning{Ezio Bartocci}
\begin{document}
\maketitle

\begin{abstract}
Decentralized monitoring 
(DM) refers to a monitoring technique, where each component must infer, based on 
a set of partial observations if the global property is satisfied.
Our work is inspired by the theoretical results presented by Baurer
and Falcone at FM 2012~\cite{Bauer201285}, where the authors introduced an 
algorithm for distributing and monitoring LTL formulae, such that 
satisfaction or violation of specifications can be detected by local monitors alone. 
However, their work is based on the main assumption that neither the computation 
nor communication take time, hence it does not take into account how to set a sampling 
time among the components such that their local traces are consistent. In this work 
we provide a timed model in UPPAAL and we show a case study on a 
networked embedded systems board.
\end{abstract}

\section{Introduction}

The majority of all computing devices produced nowadays, are embedded 
systems employed to monitor and control physical processes: cars, airplanes, 
automotive highway systems, air traffic management,
etc..  In all these scenarios, computing and communicating devices, sensors 
monitoring the physical processes and the actuators controlling the physical 
substratum are distributed and interconnected together in dedicated networks. 
In order to verify the correct behavior of these systems at runtime, 
the user often needs to monitor the emergent behavior of 
these autonomous  systems perceiving them as monolithic system, where the 
global behavior is the result of all the local behaviors.  The property to be observed is usually
specified in terms of linear-time temporal logic~\cite{Pnueli1977} (LTL) formulae or as a finite state
machine accepting the language of all the traces satisfying the property of interest.
The observation of the system can follow two different approaches. The first is the 
 \emph{centralized} observation,  where all the events generated by the local 
 components (i.e. sensors values) must be sent to a central dedicated component 
 that collects the local traces, orders them in a global trace and monitors
  the property of interest. In many real-world applications, 
 where both the communication and the number of components need to be kept minimal, this 
approach is not feasible for practical and economical reason. 
An alternative method is the \emph{decentralized} monitoring, where each components  
must infer, based on a set of partial observations if the global property is 
satisfied. Our work is inspired by the theoretical results presented by Baurer at al. at 
FM 2012~\cite{Bauer201285}, introducing  an algorithm for distributing and monitoring 
LTL  formulae, such that satisfaction or violation of specifications 
can be detected by local monitors alone.
In their paper the monitoring is performed using a technique also known as formula 
progression~\cite{Rosu2005,Bauer201285,Bacchus1998}, where the LTL formula is 
rewritten into a new formula expressing what needs to be satisfied by the current 
observation and a new formula which has to be satisfied by the trace in the future. 
In the decentralized setting, the progression is performed by each component equipped 
with a rewriting engine. In this case the monitoring may involve the exchange, 
with the other components, of messages containing the rewritten LTL formula with 
past obligations on the events not directly observable by the local component. 
Unfortunately, we found that their very elegant theoretical results are hard to 
implement in real-time embedded systems for their main assumption in which 
neither the computation nor communication take time.
For example, the sampling time with which the events are observed must be 
consistent among the components and during the monitoring.
This time depends both on the communication media, the size of the messages 
exchanged, the worst-time execution of the formula progression.
In this work we try to address these problems by providing a timed model in 
UPPAAL and we show a case study on a networked embedded 
systems board. 

The paper is organized as follows: in Section~\ref{sec:related} 
we discuss the related works and  in Section~\ref{sec:background} we
introduce the background material. In Section~\ref{sec:timedmodel} we
present a timed model in UPPAAL for sample-based monitoring and in 
Section~\ref{sec:casestudy} we show a case study. The conclusion is 
in  Section~\ref{sec:conclusion}.

\section{Related Work}\label{sec:related}

\emph{Runtime verification} (also called \emph{monitoring})~\cite{Bauer2006,Havelund2002a} 
is a lightweight yet powerful formal technique used to check whether the current 
execution of a program satisfies or violates a property of interest.  This technique 
differs from the classical and more expensive  \emph{model checking}~\cite{Clarke1982,Queille1982} 
that aims instead to verify the correctness of the property exhaustively for all the 
possible program behaviors. Monitoring is generally used when the system model is 
too big to handle with \emph{model checking} due to the state-explosion problem, or 
when the system model is not available, or it is a \emph{black-box} where only the 
 ouputs are observable. Furthermore, runtime verification can also be used to trigger 
 some system recovery actions when a safety property is violated. If the system under scrutiny is distributed, multiple 
and decentralized monitoring processes~\cite{Bauer201285,Genon2006,Sen2006,Tripakis2005,Wang,Wang2007}
 can be employed to check if during the execution a global property is satisfied or not.
In~\cite{Sen2006,Sen2004418} the authors describe an efficient decentralized 
monitoring algorithm, based on a variant of past time linear temporal logic,  that 
monitors a distributed program's execution to check for violations of safety properties. 
However, their work does not deal with time constraints and does not address 
real-time applications running on  networked embedded systems.  
Baur and Falcone propose in~\cite{Bauer201285}  an algorithm for decentralized 
LTL monitoring in synchronous systems based on formula progression/rewriting, 
but also in that work, neither the overhead of monitoring and the computation time are 
 taken into account. In the last years, several techniques have been developed to 
control the overhead~\cite{Callanan2008,Stoller2011,Bartocci2012,Kalajdzic2013} of monitoring.
The majority of these techniques involve the use of \emph{event-triggered} monitors, 
where the monitor is invoked whenever a new event is triggered by the system, 
making the overhead unpredictable.
Our approach is based on  \emph{synchronous sampling} and 
it is similar to the one introduced in~\cite{Bonakdarpour2011},
but extended for the case of networked embedded systems, where the local 
monitors take samples from the local program variables and the sensors to 
analyze if a property of interest is satisfied or not.

\section{Background}\label{sec:background}

\begin{figure}[h!]
  \centering
      \includegraphics[width=0.95\textwidth]{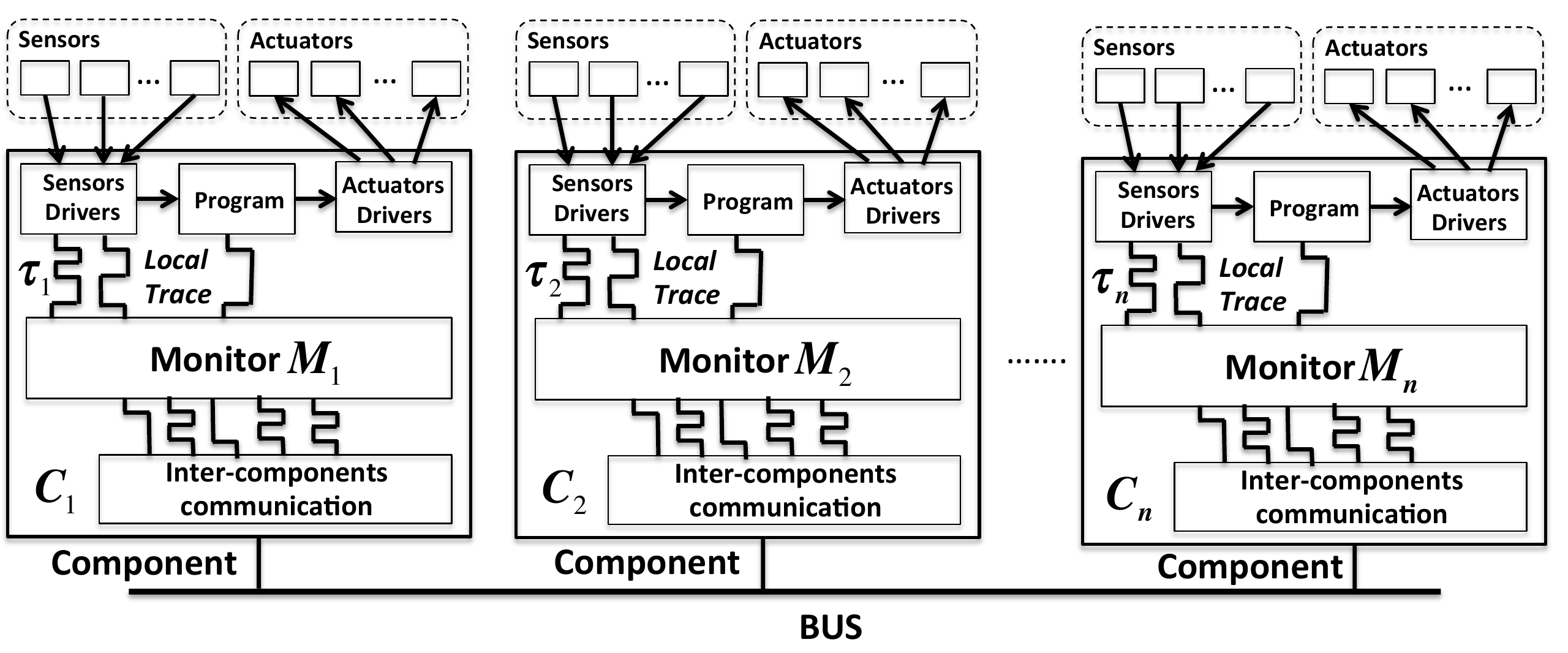}
   \vspace{-1mm}
  \caption{Decentralized monitoring setting for a distributed embedded systems.}
  \label{fig:decentralized_setting}
\end{figure}

\noindent In the \emph{decentralized monitoring} setting considered in this paper, 
we assume a set $\mathcal{C}=\{C_1,C_2, \cdots, C_n\}$  of components 
communicating on a serial communication BUS as Fig.~\ref{fig:decentralized_setting} shows.  
Each component $C_i$ is equipped with a monitor $M_i$ that can observe the set of events $\Sigma_i$. 
$\Sigma = \Sigma_1 \cup \Sigma_2 \cup \cdots \cup \Sigma_n$, is the set of
all events. If each event can be observed only by a single component, we assume 
that $\Sigma_i \cap \Sigma_j = \emptyset$ for all $i,j \leq n$ with $i \neq j$.
The property to be monitored  is specified in a LTL~\cite{Pnueli1977} formula $\varphi$ 
over a set of propositions AP, such that $\Sigma=2^{AP}$. We denote
with $\tau_i(m)$ the (m+1)-th event in the \emph{local trace} $\tau_i$ 
observed by the monitor $M_i$ and with $\mathbf{\tau} = (\tau_1, \tau_2, \cdots, \tau_n)$ 
the \emph{global trace} such that $\tau = \tau_1(0) \cup \tau_2(0) \cup \cdots 
\tau_n (0) \cdots \tau_1(t) \cup \tau_2(t) \cup \cdots \tau_n (t)$.  Each component is an
embedded computing device that can access the values of a set of external sensors
 (i.e. the external temperature or the button pressure) and can control through a 
program a set of actuators (i. e. fan speed or the temperature of the heater). 
Each monitor observes the change of the local sensors and the program 
values together with the events that may receive from the other components
through a serial communication BUS, enabling the exchange of 
multicast messages.

\begin{defn}[Linear Temporal Logic (LTL) Syntax~\cite{Pnueli1977} ] The syntax for an 
LTL formula is described by the following grammar:
$$\varphi::= \top \mbox{ } |  \bot \mbox{ } | p  \mbox{ } |  \mbox{ } \neg \varphi  \mbox{ } | 
\mbox{ } \varphi \wedge \varphi \mbox{ } |  \mbox{ } \mathbf{X} \varphi | \mbox{ } \varphi \mathbf{U} \varphi$$

\noindent where $p \in AP$. A LTL formula has atomic propositions $p$, 
logical connectives $\neg$ $\wedge$, temporal operators $\mathbf{X}$ (next), $\mathbf{U}$ (until).
As usual, we introduce shorthands
by defining the following derivative logical and temporal operators:
$$
\begin{array}{*{20}l}
\begin{array}{l}
\psi \vee \varphi \\
\psi \rightarrow \varphi \\
\psi \leftrightarrow \varphi \\
\Box \psi \\
\Diamond \psi\\
\end{array}
\begin{array}{c}
\Leftrightarrow \\
\Leftrightarrow \\
\Leftrightarrow \\
\Leftrightarrow \\
\Leftrightarrow \\
\end{array}
\begin{array}{l}
 \neg (\neg \psi \wedge \neg \varphi) \\
 \neg \psi \vee\varphi \\
 (\psi \rightarrow \varphi) \wedge (\varphi \rightarrow \psi) \\
 \neg (\bot\mbox{ } \mathbf{U}\neg \psi)\\
 \top \mbox{ } \mathbf{U} \psi\\
 \end{array}
\begin{array}{l}
\mbox{  or } (\vee)\\
\mbox{  implies } (\rightarrow)\\
\mbox{  equivalent } (\leftrightarrow) \\
\mbox{  always } (\Box)\\
\mbox{  eventually } (\Diamond) \\
\end{array}
\end{array}
$$
\end{defn}

\begin{defn}[Linear Temporal Logic (LTL) Semantics~\cite{Pnueli1977} ] Let be
$\tau = \tau_1(0) \cup \tau_2(0) \cup \cdots 
\tau_n (0) \cdots \tau_1(t) \cup \tau_2(t) \cup \cdots \tau_n (t) \cdots \in \Sigma^{\omega}$ (global trace) an infinite word with $i \in \mathbb{N}$ being 
a position corresponding to a particular time step. Then the semantics of an LTL formula is defined inductively as follows:

$$
\begin{array}{*{20}c}
   \begin{array}{l}
    \tau, i  \models \top \mbox{ holds } \\
    \tau, i  \models \bot \\
    \tau, i  \models p \\
      \tau, i  \models \neg \varphi \\
    \tau, i  \models \varphi_1 \wedge \varphi_2 \\
    \tau, i  \models \mathbf{X} \varphi \\
     \tau, i  \models \varphi_1 \mathbf{U} \varphi_2 \\

  \end{array} &
     \begin{array}{c}
     \mbox{ (is true).}\\
       \Leftrightarrow \\
   \Leftrightarrow \\
  \Leftrightarrow \\
  \Leftrightarrow \\
 \Leftrightarrow \\
   \Leftrightarrow \\
  \end{array}   &
       \begin{array}{l}
       \\
   \tau, i  \not \models \top \\
    p \in \tau_1(i) \cup \tau_2(i) \cup \cdots \tau_n (i)\\
    \tau, i \not  \models \varphi \\
     \tau, i \models \varphi_1 \mbox{ and } \tau, i \models \varphi_2 \\
\tau, i+1  \models \varphi \\
     \exists k \geq i \mbox{ } \tau, k \models \varphi_2 \mbox{ and } \forall i \leq l < k \mbox{ }  \tau, l \models \varphi_1\\
  \end{array}
 \end{array}
$$
Moreover, $\tau  \models \varphi$  holds  $ \Leftrightarrow$ $\tau,0  \models \varphi$.

\end{defn}
We denote $\mathcal{L}(\varphi) = \{w \in \Sigma^{\omega} | w \models \varphi\}$ as the language 
generated by an LTL-formula $\varphi$ and corresponding to a set of models of a LTL-formula
$\varphi$. The languages generated by two formulae $\varphi$ and $\psi$ are the same
$\mathcal{L}(\psi)=\mathcal{L}(\varphi)$ iff $\varphi \equiv \psi $. A common technique to 
verify the correctness of a property is to generate a monitor from a LTL-formula. 
Such a monitor can be then executed in parallel with the application 
to be verified at runtime (\emph{online synchronous monitoring}) or can be 
used after the program execution to check a finite set of recorded executions (\emph{offline monitoring}). 
There are two main approaches to generate synchronous monitors.
The first method relies on the generation of \emph{automata-based} monitors. 
In particular, there are several papers~\cite{Etessami2000,Gastin2003,Wolper2000} describing how to build
 a reduced nondeterministic B\"uchi automaton~\cite{Buchi1962} able 
to recognize infinite words of the language $\mathcal{L}(\varphi)$ of a LTL formula $\varphi$. A B\"uchi automaton
can be then turned to a monitor~\cite{damorin2005,Geilen2001}  in the form of a deterministic 
finite state machine (DFSM). Generally, the process of converting 
a LTL formula into a monitor is expensive and the size of the  B\"uchi automata generated can be
$2^{O(|\varphi|)}$~\cite{Gastin2003}. However, once the monitor is generated,
its execution can be very efficient. In particular, Rosu et. at. showed
in~\cite{damorin2005} how to build particular DFSMs called  
\emph{binary transition tree finite state machines}  (BTT-FSM) that 
perform a transition from a state to another state of the monitor by 
evaluating an optimal number of atomic propositions.  

\begin{figure}[ht!]
  \centering
      \includegraphics[width=0.95\textwidth]{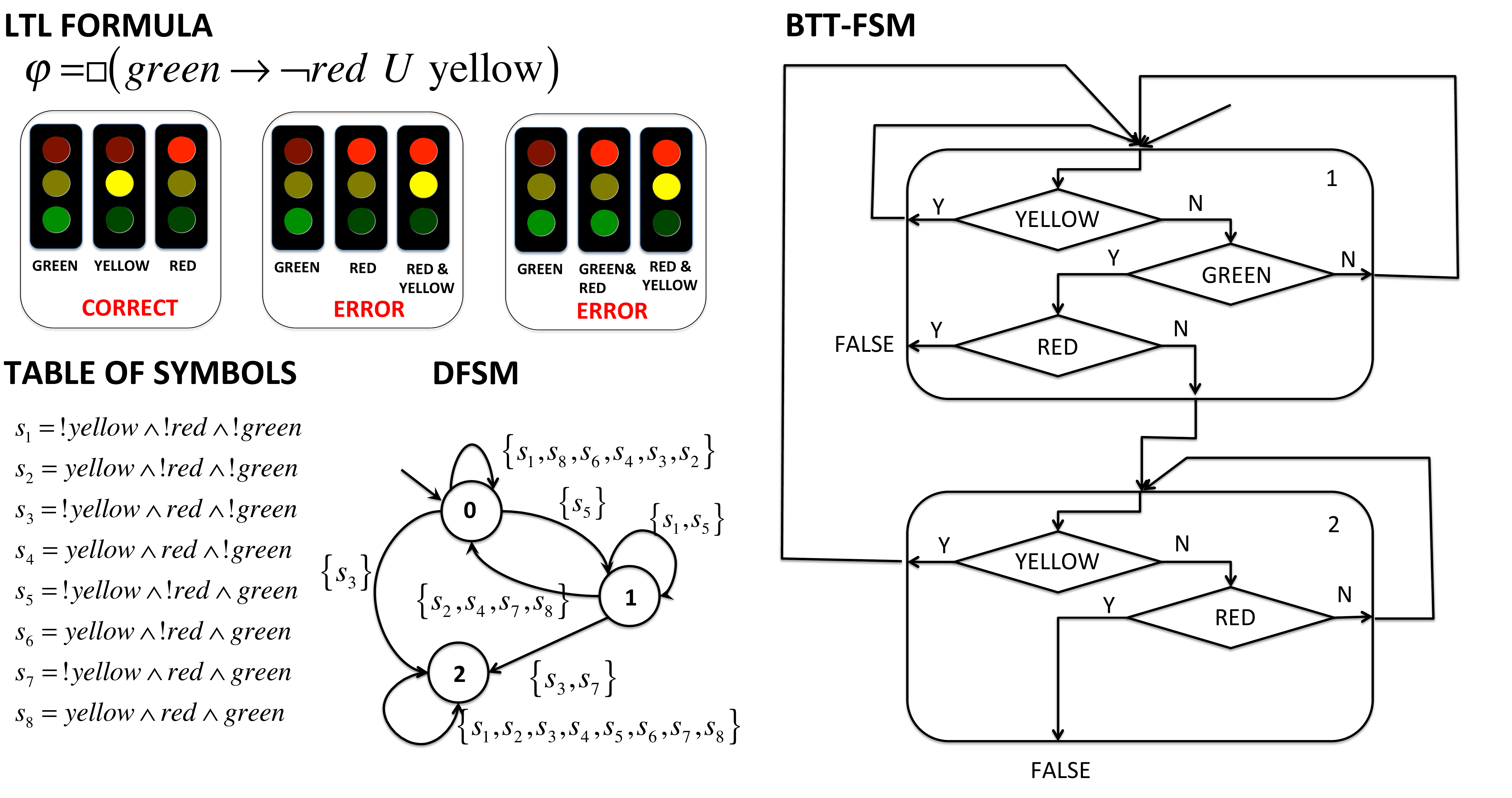}
      
   \vspace{-2mm}
  \caption{Automata-based monitors to check the property of a traffic light that always when it is green, then it is not red until it is yellow.}
  \label{fig:figure_monitors}
\end{figure}

\noindent An alternative monitoring approach is 
based on \emph{formula rewriting}~\cite{Rosu2005,Bauer201285}
 or \emph{formula progression}~\cite{Bacchus1998}. 
The monitor in this case is a rewriting engine, that rewrites 
the current formula into a new formula expressing what 
needs to be satisfied by the current observed events and 
what are the future obligations to meet. The overhead 
required for monitoring with this approach is 
higher than by using a DFSM. On the other hand this 
method is more flexible, because does not require 
a process of translation from LTL formula to monitor
and allows to change at runtime the formula to be monitored.  In the following, 
we provide some basic definitions for the LTL rewriting function and
the monitoring result.
 
 \begin{defn}[LTL rewriting function\cite{Bauer201285}] Let $S_{LTL}$ be the set of all the possible LTL formulae 
 and $\phi, \phi_1, \phi_2 \in S_{LTL}$, $\sigma \in \Sigma$ an event, the LTL rewriting function 
 $R:S_{LTL} \times \Sigma \rightarrow S_{LTL}$ is inductively defined as follows:
 
$$
\begin{array}{*{20}c}
   \begin{array}{rcl}
      R(\top, \sigma) & = & \top \\ 
      R(\bot, \sigma) & = & \bot \\ 
      R(\neg \phi, \sigma)            & = & \neg R(\phi,\sigma) \\
      R(\mathbf{X} \phi, \sigma)   & = & \phi \\
  \end{array} &
     \begin{array}{rcl}
       R(p \in AP,\sigma) & = & \top \mbox{, if } p \in \sigma, \bot \mbox{ otherwise}  \\ 
       R(\phi_1 \vee \phi_2, \sigma) & = & R(\phi_1,\sigma) \vee R(\phi_2,\sigma)\\ 
       R(\phi_1 \mathbf{U} \phi_2, \sigma) & = & R(\phi_2,\sigma) \vee R(\phi_1,\sigma) \wedge \phi_1 \mathbf{U} \phi_2\\
       R (\square \phi, \sigma) & = & R (\phi, \sigma) \wedge \square \phi  \\
       R (\diamond \phi, \sigma) & = & R (\phi, \sigma) \vee \diamond \phi  \\
  \end{array} 
 \end{array}
$$

 \end{defn}
  
\begin{defn}[Monitoring\cite{Bauer201285}]  Let $u \in \Sigma^{*}$ denote a finite word. The evaluation  
of the \emph{satisfaction relation}, $\models_{3}:\Sigma^{*} \times LTL \rightarrow \mathbb{B}_3$, with
$\mathbb{B}_3 = \{\top, \bot, ?\}$  of a formula $\varphi$ with respect to $u$ is defined as:
$$ u \models_{3} \varphi = \left\{ \begin{array}{lcl} 
                                           \top & \mbox{ if } & \forall \sigma \in \Sigma^{w}: u\sigma \models \varphi\\
                                           \bot & \mbox{ if } & \forall \sigma \in \Sigma^{w}: u\sigma \not\models \varphi\\
                                              ?   & \mbox{otherwise}\\
                                        \end{array}\right. $$

 \end{defn}

 \section{A Timed Model for Decentralized Monitoring}\label{sec:timedmodel}
 
In this section, we propose a timed model for the decentralized monitoring  
 using networks of timed automata~\cite{Alur1994}. This formal specification allows 
us to analyze, with tools like UPPAAL~\cite{Behrmann2006}, the timing 
behavior of the system and to check important properties such as the 
synchronization of the sampling, the sampling time and granularity.  
A timed automaton is a finite-state machine enriched with clock 
variables using a dense-time model. For the sake of completeness, in the 
following we provide all necessary definitions.

 \begin{defn}[Timed Automaton (TA)~\cite{Behrmann2006}] A timed automaton is a tuple 
$\mathcal{A} = (L, l_0, C, \Sigma, E, I)$ where:

\begin{itemize}
\item $L$ is a finite set of locations,
\item $l_0 \in L$ is the initial location,
\item $C$ is a finite set called the clocks of $\mathcal{A}$,
\item $\Sigma$ is a finite set called the alphabet or actions of $\mathcal{A}$,

\item $E \subseteq L \times \Sigma \times B(C) \times 2^C \times L$ is a set of edges, called transitions of $\mathcal{A}$, where
 B(C) is the set of conjunctions over simple conditions of the form $x \bowtie c$ or 
                      $x - y \bowtie c$, where $x,y \in C$, $c \in \mathbb{N}$ and $\bowtie \in \{<, \leq, =, \geq, >\}$,
\item $I: L \rightarrow B(C)$ assigns invariants to locations.
\end{itemize}
 \end{defn}
 
  \begin{defn}[Semantics of TA\cite{Behrmann2006}]  Let $\mathcal{A} = (L, l_0, C, \Sigma, E, I)$ 
  be a timed automaton. The semantics is defined as a labelled transition system 
  $\langle S, s_0, \rightarrow \rangle$, where: 
  
  \begin{itemize}
 \item  $S \subseteq L \times \mathbb{R}^C$ is the set of states, 
 \item $s_0 = (l_0, u_0)$ is the initial state,
 \item $\rightarrow \subseteq S \times (\mathbb{R}_{\geq 0} \cup \Sigma) \times S$
 is the transition relation such that:
      \begin{itemize}
               \item $(l,u)$ $\xrightarrow{d}$ $(l, u + d)$ if $\forall d': 0 \leq d' \leq d$ $\Longrightarrow$ $u+d' \in I(l)$, 
               \item $(l,u)$ $\xrightarrow{a}$ (l',u') if there exists $e=(l,\sigma,g,r,l') \in E$ s.t. $u \in g$,
               $u' = [ r \longmapsto 0]u$, and $u' \in I(l')$, 
       \end{itemize}
       where for $d \in \mathbb{R}_{\geq 0}$, $u + d$ maps each clock $x$ in $C$ to the value $u(x) + d$, and 
      $ [r \longmapsto 0]u$ denotes the clock valuation which maps each clock in $r$ to 0 and agrees with $u$ over $C \setminus r$.
 \end{itemize}
 \end{defn}
A network of timed automata~\cite{Behrmann2006} is defined as a parallel composition of timed automata 
over a common set of clocks and actions, consisting of $n$ timed automata 
$\mathcal{A}_i = (L_i, l^0_i, C, \Sigma, E_i. I_i), 1 \leq i \leq n$.  A location vector is a vector 
$\overline{l} = (l_1, \cdots, l_n)$. The invariant functions are composed in a common 
function over location vectors $\overline{l} = \wedge_i  I_i(l_i)$. Following the notation in~\cite{Behrmann2006}, 
we denote with $\overline{l}[l_i'/l_i]$ the vector where the $i$th element $l_i$ of $\overline{l}$ 
is replaced by $l'_i$. 

  \begin{defn}[Semantics of a network of Timed Automata~\cite{Behrmann2006}] 
  Let $\mathcal{A}_i = (L_i,l^0_i, C, \Sigma, E_i, I_i)$ be a network of $n$ timed automata. 
  Let $\overline{l}_0 = (l_1^0, \cdots, l_n^0)$ be the initial location vector. The semantics is 
  defined as a transition system $\langle S, s_0, \rangle$, where $S = (L_1 \times \cdots \times L_n) \times \mathbb{R}^C$ is 
  the set of states, $s_0 = (\overline{l}_0, u_0)$ is the initial state, and 
  $\rightarrow \subseteq S \times S$ is the transition relation defined by:
  \begin{itemize}
  \item $(\overline{l}, u) \xrightarrow{d} (\overline{l}, u + d)$ if $\forall d':0 \leq d' \leq d \Longrightarrow u + d' \in I(\overline{l})$.
  \item $(\overline{l}, u) \xrightarrow{a} (\overline{l}[l_i'/l_i],u')$ if there exists $l_i  \xrightarrow{\tau g r} l_i'$ s.t. $u \in g$,
          $u' = [r \longmapsto 0]u$ and $u' \in I(\overline{l}[l_i'/l_i])$.
   \item $(\overline{l}, u) \underrightarrow{a} (\overline{l}[l_j'/l_j, l_i'/l_i], u')$ if there exist $l_i  \xrightarrow{c?g_ir_i} l_i' $ and\\
   $l_j  \xrightarrow{c!g_j r_j} l_j'$ s.t. $u \in (g_i \wedge g_j)$, $u' = [r_i \cup r_j  \longmapsto 0]u$ and $u' \in I(\overline{l}[l_j'/l_j, l_i'/l_i])$. 
  \end{itemize}

 \end{defn}

\begin{figure}[ht!]
  \centering
      \includegraphics[scale=0.4, bb=800 0 200 800]{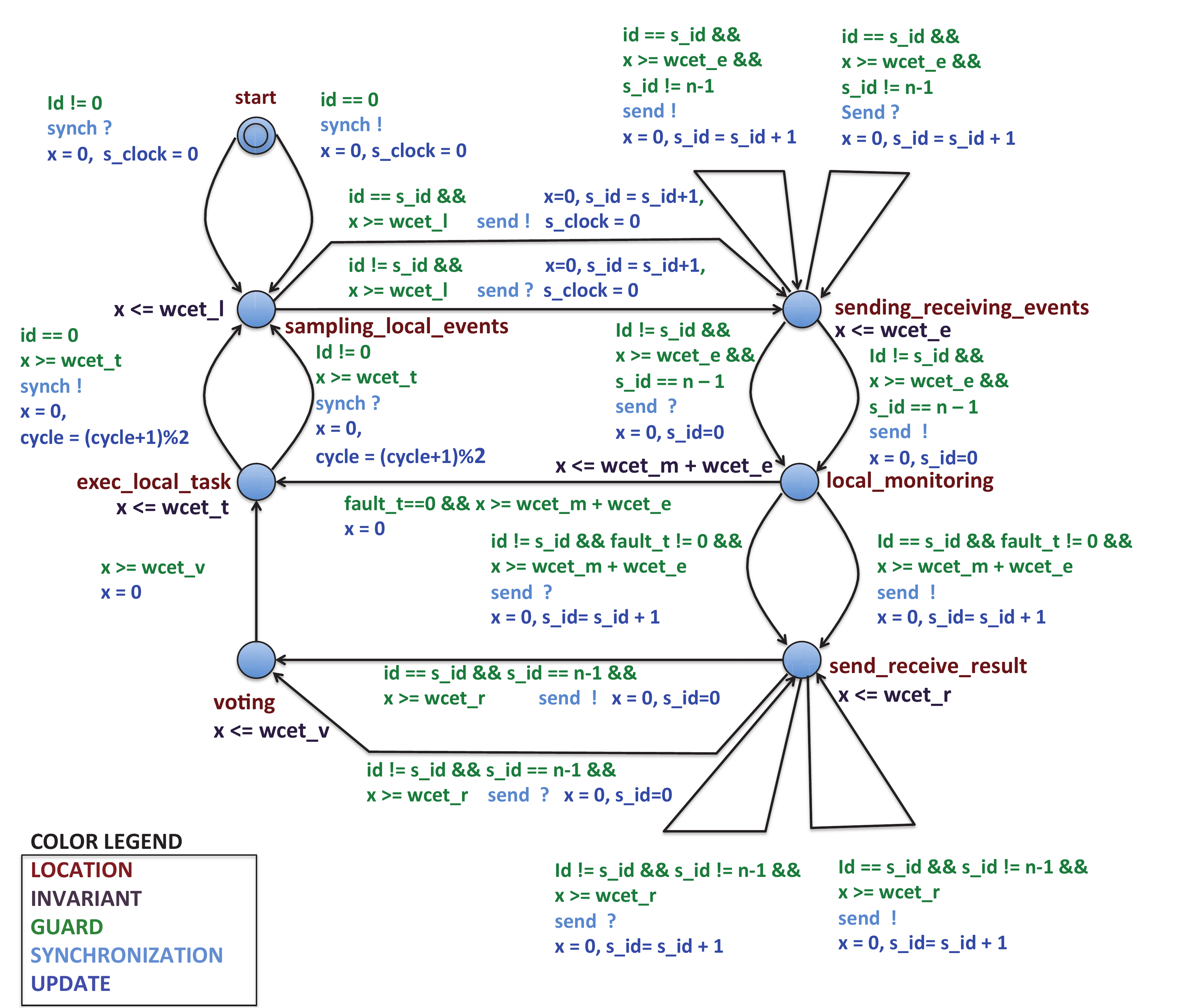}
   \vspace{-1mm}
  \caption{TA specification for each component involved in the decentralized monitoring.}
  \label{fig:timed_model}
\end{figure}
The UPPAAL standard semantics presented in Definitions 4.1, 4.2 and 4.3 includes neither the use 
of \emph{bounded integer variables}, nor the use of \emph{broadcast channels}. Variables
allow to keep low the number of locations to handle, while the semantics of the broadcast
channel  does not require to have receivers synchronized and so is never blocking.
In our timed-model we employ both of these UPPAAL extensions and we refer 
 the reader to~\cite{Behrmann2006} for further details. 

     Fig.~\ref{fig:timed_model} shows the timed model\footnote{The UPPAAL model can be downloaded at 
\url{www.eziobartocci.com/has/decentralized_monitoring.xml}}  
chosen for each component $C_i$ in Fig.~\ref{fig:decentralized_setting}.  The model provides two 
different possible behaviors depending 
on the value $\{0,1\}$ of the parameter  $fault_t$. This parameter enables/disables a fault-tolerance 
mechanism  called N modular redundancy (NMR) in which $2 k + 1$ modules perform a process (in this case the monitoring) 
and the result is processed by a voting system to produce a single output.  For example, 
in triple modular redundancy (TMR) if any one of the three systems fails, the other two systems 
can correct and mask the fault. The other important parameters in the model are the worst-case execution 
time  (WCET) of the tasks involved in the process. WCET measures the maximum time length a task 
could take to execute on a specific hardware platform. In our setting we consider the following parameters:
\begin{itemize}
\item $wcet_l$ is the WCET to sample all the new local events to be monitored, 
\item  $wcet_e$ is the WCET to send a message with the changed events from one node
to the others (note that the communication is multicast), 
\item $wcet_m$ is the WCET to monitor the events,
\item $wcet_r$ is the WCET of sending a message with the result from one node to the others,
\item $wcet_v$is the WCET to perform the voting,
\item $wcet_t$ is the WCET to execute a local task
\end{itemize}

\noindent Measuring the WCET is in the general case insoluble, because it is equivalent to the \emph{halting problem}. 
However, in many particular  cases (i.e. when the software does not contains infinite loops) is still 
possible to provide an over-approximation of such measure. The most common techniques to 
calculate the WCET are static analysis (by reasoning on the control graph, without executing the code) 
of the software or by runtime measuring the performances  through the generation of appropriate test cases.
\noindent All the WCET parameters are used to determine how much time each component 
$C_i$ should stay in a particular location described in the timed model of Fig. \ref{fig:timed_model}.
A clock $x$ is used to keep track of the elapsed time in a location. This clock variable is 
reset when an enabled transition (representing an action) is taken and it is constrained with 
one of the WCET parameters mentioned before to determine the max time allowed in 
a particular location. Another clock variable $s_{clock}$ keeps track of the 
time length elapsed between one sampling and the next one and it is used later to 
perform the analysis through model checking.

\begin{figure}[htbp]
  \centering
  \includegraphics[width=70mm]{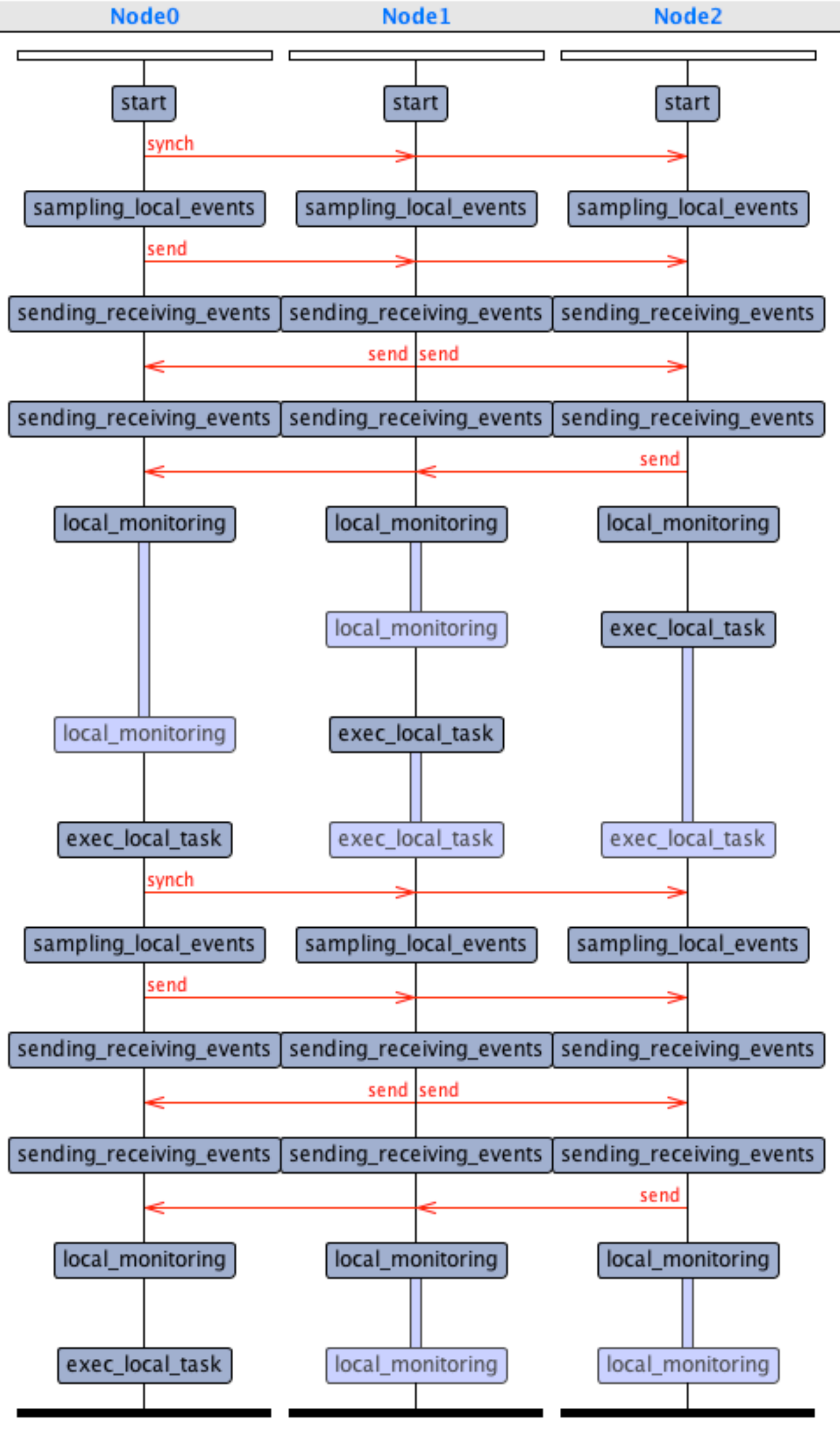}%
  \qquad
  \includegraphics[width=65mm]{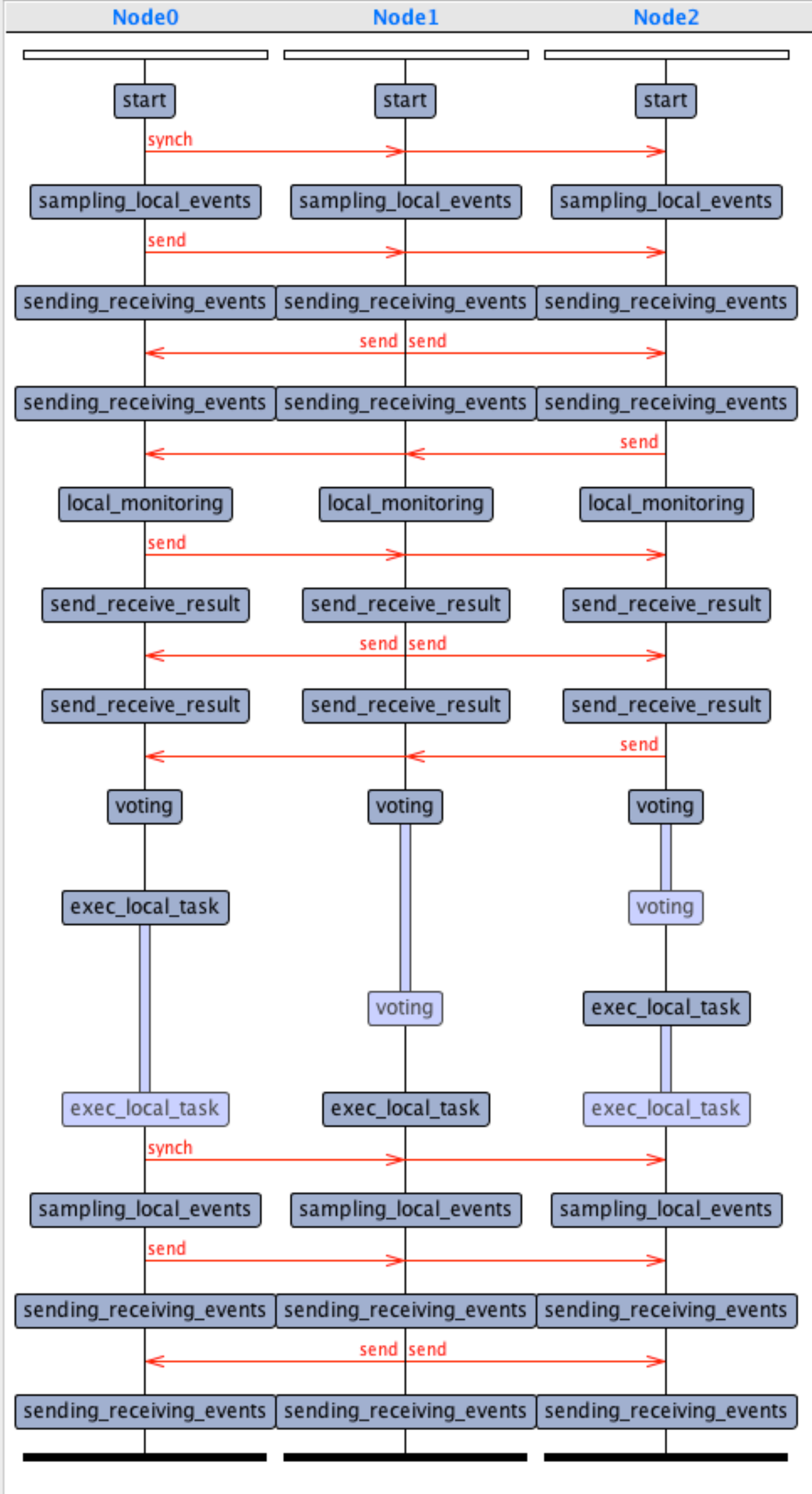}
  \caption{Two possible sequence (on the left $fault_t$=0, while on the right $fault_t$=1) diagrams 
  generated using UPPAAL simulator for the model in Fig. \ref{fig:timed_model}. } 
  \label{fig:sequence_diagram}
\end{figure}
\noindent Two broadcast channels $synch$ and $send$ are used to realize the multicast 
communication.  In a broadcast synchronization one sender with the actions $synch !$ or $send !$ 
can synchronize with an arbitrary number of receivers through the action $synch ?$ or $send ?$.  
If a receiver in its current state has an enabled transition in which it can synchronize, it must do so. 
However, the broadcast sending is never blocking, so the sender can execute a synchronization 
action even if there are no receivers. 
 A variable $cycle$ is used for analysis purposes to mark the current cycle from the next and 
 the previous one.  Fig. \ref{fig:sequence_diagram} shows two possible execution 
 traces of the system, one with the NMR enabled and one with NMR disabled. The timed 
 automaton starts (\emph{start} location) with a synchronization action \emph{synch} sent 
 always by the first component with $id==0$ and received by the other components 
 with $id != 0$, respectively.  Then in the \emph{sampling\_local\_events} location  
within $wcet_l$ milliseconds all the 
local events in each nodes are sampled. The events that are changed in each node, are 
sent (in the location \emph{sending\_receiving\_events}) to the others with $n$ multicast messages, where $n$ is the number of nodes.
The order with which the nodes exchange their messages follows the order of their $id$ 
(i.e. the node with lower $id$ starts first). In the location \emph{local\_monitoring}, 
the local monitor processes the events and produces a result $\{\bot, \top, ?\}$. 
If the fault-tolerant mechanism is enabled ($fault_t$==1), the result is sent from each node 
to all the other nodes (in \emph{send\_receive\_result}) and a voting mechanism will 
follow (\emph{voting} location). A local task (i.e. displaying results, increase the 
heater temperature, etc..) can also be executed in the location \emph{exec\_local\_task}, 
before the local sampling will start again the loop.
In the following we shows some properties that are possible to be verified in the proposed 
timed model using UPPAAL tool.
\begin{prop} [Liveness] When the timed automaton in Fig. \ref{fig:timed_model} will enter the sampling\_local\_events 
location at the $cycle=i$ will then eventually enter the same location  at $cycle=(i + 1) \mbox{ mod } 2$
with $i \in \{0,1\}$. 
\end{prop} 


\noindent In UPPAAL this property can be expressed using the \emph{leads to} or
\emph{response} form, written $\varphi \leadsto \psi$ which means whenever $\varphi$ 
is satisfied, then $\psi$ will be satisfied. It is possible to verify that the following liveness
property 
$((\mathcal{C}_i.\mbox{sampling\_local\_events} \wedge cycle==0) \leadsto (\mathcal{C}_i.\mbox{sampling\_local\_events} \wedge cycle==1))
\wedge ((\mathcal{C}_i.\mbox{sampling\_local\_events} \wedge cycle==1) \leadsto (\mathcal{C}_i.\mbox{sampling\_local\_events} \wedge cycle==0))$
holds for each $1 \leq i \leq n $.

\begin{prop} [Synchronous sampling] Given a network of $n$ timed automata, there is not a 
reachable state, where one timed automaton is in the sampling\_local\_events location and the others in 
different locations at the same time. This means that the local events will be sampled by each 
component always synchronously.
\end{prop} 
This property can be expressed in UPPAAL as the negation of a path formula (\emph{exist eventually}) $\neg E \Diamond \varphi$:
$\neg E \Diamond ( \bigvee_{1 \leq i \leq n} (\mathcal{C}_i.\mbox{sampling\_local\_events } \wedge   (\bigvee_{j \neq i, 1 \leq j \leq n} \neg \mathcal{C}_j.\mbox{sampling\_local\_events }  ) ))$

\begin{proposition} [Sampling frequency]
Given a network of $n$ components with the timed model shown in Fig. \ref{fig:timed_model}, 
the sampling frequency function $f_{s}:\mathbb{N} \rightarrow \mathbb{R}$ with which the 
local events are sampled is:

$$f_{s}(n) = \frac{1}{wcet_l + n \cdot  wcet_e + wcet_m + fault_t \cdot (n \cdot wcet_r + wcet_v) + wcet_t}$$
 with $wcet_l$ time units is also the time granularity and within this interval of time it is not possible to 
distinguish two different samples.
\end{proposition}

\noindent In UPPAAL we can check that the sampling period is always constant, by verifying the following formula:
$(\mathcal{C}_i.s_{clock} == 0)  \leadsto    (\mathcal{C}_i.sampling\_local\_events \wedge \mathcal{C}_i.s_{clock} == wcet_l + n \cdot  wcet_e + wcet_m + fault_t \cdot (n \cdot wcet_r + wcet_v) + wcet_t)$

We verified the previous properties in UPPAAL by varying the number of components $n$ from two to ten. However,
we can generalize to an arbitrary number of nodes by making the following observations on 
the timed-model of  Fig. \ref{fig:sequence_diagram} (here we consider only the case $fault_t=0$, but the observations are similar for the case  $fault_t=1$):
\begin{enumerate}
\item \textbf{Start} $\rightarrow$ \textbf{sampling\_local\_events}. All $n$ components are initialized in 
the \emph{Start} location. The first transition is forced by 
the component $\mathcal{C}_0$ that synchronizes with a $synch\mbox{ }!$ action all the other components $\mathcal{C}_i$ 
with $0 < i < n$ to switch, with a $synch\mbox{ }?$ action (the only one enabled), into the new location 
\emph{sampling\_local\_events} at the same time. There 
is no possibility that one component is in location  \emph{Start} and another is in location 
\emph{sampling\_local\_events}.

\item \textbf{sampling\_local\_events}$\rightarrow$ \textbf{sending\_receiving\_events}. In this case 
all the components need to wait the same amount of time $wcet_l$ even if one finishes to sample 
the local events before another. After  $wcet_l$ time there is only one action enabled for each component: 
 $send\mbox{ }!$ for  $\mathcal{C}_0$ and  $send\mbox{ }?$ for  $\mathcal{C}_i$ with 
 $0 < i < n$. This step models $\mathcal{C}_0$ sending its local events update to 
 all the other components.

\item \textbf{sending\_receiving\_events}$\rightarrow$ \textbf{sending\_receiving\_events}. 
A sequence  of $n-2$ broadcast synchronization actions $send\mbox{ }!$ and $send\mbox{ }?$
will be enabled after waiting $wcet_e$ time each step. Incrementing the variable $s_{id}$
from one to $n-2$ will distinguish the sender component  $\mathcal{C}_{i=s_{id}}$ performing 
the synchronization action $send\mbox{ }!$ from the receiver components $\mathcal{C}_{i \neq s_{id}}$ 
performing  the action $send\mbox{ }?$. This sequence of synchronizations costs $(n-2) \cdot  wcet_e$
time.

\item \textbf{sending\_receiving\_events}$\rightarrow$ \textbf{local\_monitoring}. This transition is 
enabled only when $s_{id}=n-1$ and corresponds to the last component  $\mathcal{C}_{i=n-1}$
sending its local events update to the other components after all having waited $wcet_e$ time.  

\item  \textbf{local\_monitoring}$\rightarrow$ \textbf{exec\_local\_tasks}. This transition is performed 
by all the components without synchronization. Each component should wait in the location 
 \emph{local\_monitoring} exactly  $wcet_m + wcet_e$ time and then switch to the location
 \emph{exec\_local\_tasks}.
 
\item  \textbf{exec\_local\_tasks}$\rightarrow$ \textbf{sampling\_local\_events}. This 
case is similar to 1. The time spent in the location \emph{exec\_local\_tasks} for each component is 
$wcet_t$. This transition makes sure that the \emph{liveness} Property 4.4 holds.
\end{enumerate} 

\noindent By summing up the times spent in each location for a complete cycle, it is easy to show that also 
the Proposition 4.6 holds.

 \section{Case Study}\label{sec:casestudy}

\begin{figure}[h!]
  \centering
      \includegraphics[  width=0.95\textwidth]{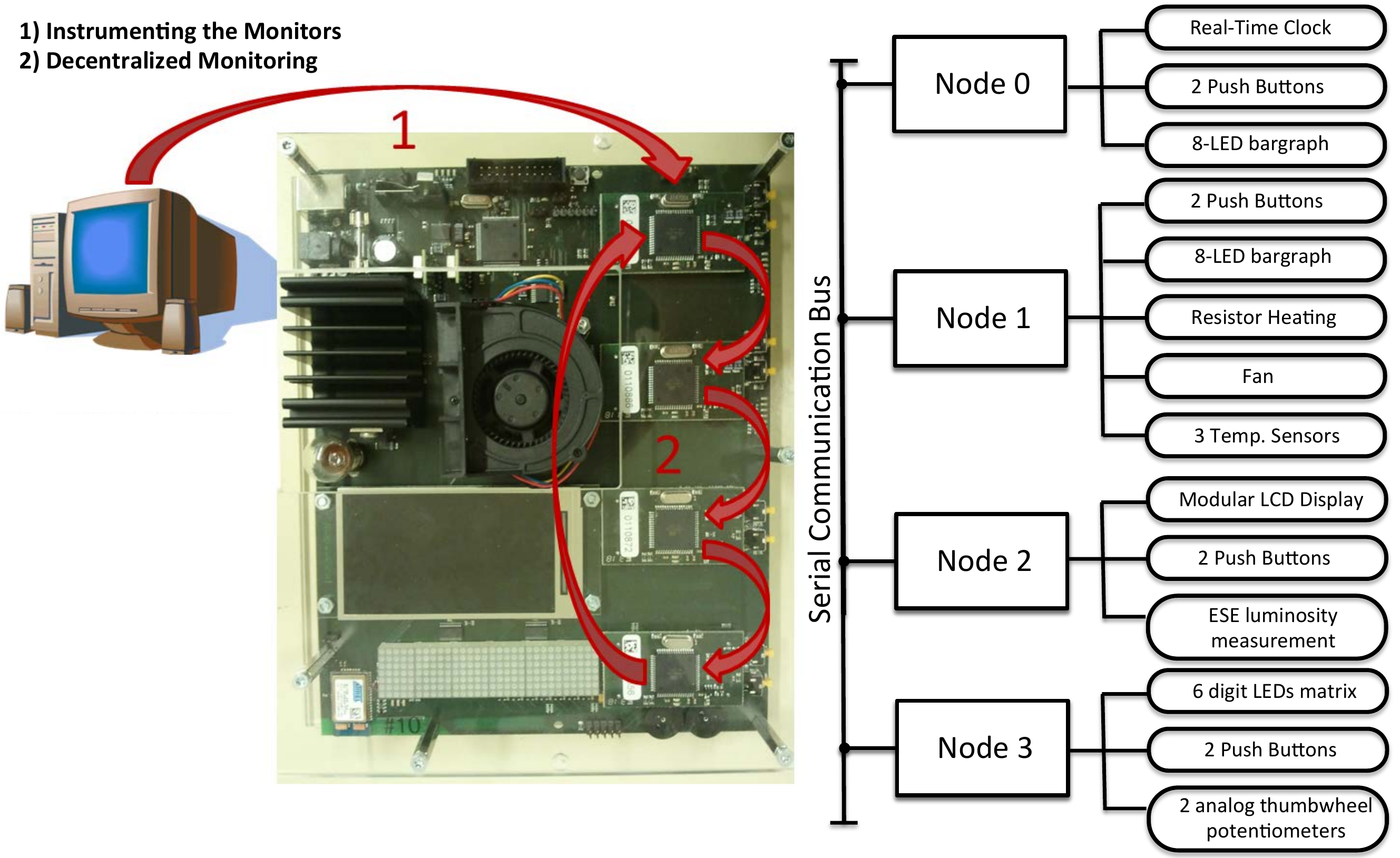}
    
   \vspace{-1mm}
  \caption{Networked Embedded Systems.}
  \label{fig:hardware}
\end{figure}

The model presented in the previous section has been implemented on 
an hardware platform (designed in our lab) hosts with four 
independent micro-controllers (ATMega128 produced by Amtel) 
nodes connected to a real-time network. Each node is equipped 
with different peripheral devices as Fig.~\ref{fig:hardware} shows. 
A shared communication BUS is included for Real-Time data transfer 
between the four nodes.

\noindent We chose the Carrier Sense Multiple Access (CSMA) with collision 
detection as our low level communication protocol among the nodes. This 
avoids that simultaneous messages are sent from the nodes and the messages do not 
require a fixed message length like in other protocols (like TTP~\cite{Kopetz2001}).
It is possible to determine the WCET of the communication 
by analyzing the max length of the exchanged messages as Table~\ref{tbl:wcet_communication}
shows. The max length of the messages depends usually on the max number of 
 atomic propositions that can change at runtime in one component.

\begin{table}
\begin{center}
    \begin{tabular}{ | l | r |  p{7cm} |}
    \hline
    \textbf{Property} & \textbf{Value} & \textbf{Description} \\ \hline
    max length of the token msg & 66 bytes &  1 byte for CSMA + 1 byte for the message length (4 bits) and token (4 bits) + 64 bytes for the data. \\ \hline
    max length of the result msg   &  4 bytes &  1 CSMA byte + 1 byte for the message length (4 bits) and token (4 bits) + 2 bytes for the data.\\ \hline
    max length of the synch msg          &   1 bytes & 1 CSMA byte \\ \hline
    max num. bytes sent in one round          &   281 bytes & synch + 4 * token msg + 4 * result msg \\ \hline
    max bits sent in one round                        & 2810 bits & a byte sent contains 8 data bits, 1 start bit  and 1 bit stop \\ \hline
   Baud rate                                                     & 4800 bit/s & \\ \hline
   Worst Case Time for communication           & 0.585 sec     & \\ \hline
    \end{tabular}
     \caption{The Worst Case execution depends on the number of micro-controllers used, the max length of a token message exchanged, 
    the max length of the result message and the baud rate of the BUS. }
\label{tbl:wcet_communication}
\end{center}
\end{table}

\noindent We have adopted both the \emph{automata-based} and the \emph{formula 
progression monitoring} approaches.  In the automata-based approach,
we have used  LTL3 tools\footnote{\url{http://ltl3tools.sourceforge.net}} to generate the DFSM from a LTL formula  
and then coded the resulted 
state-machine in C, while to measure the monitoring overhead is possible to use a 
static analyzer for AMTEL micro-controller like Bound-T\footnote{\url{http://www.bound-t.com}} . 
Concerning the formula-progression monitoring technique, even if we have imposed some 
limitations on the length of the formula (max 64 symbols) and on the number of next 
temporal operators allowed, the only way to measure the WCET is by 
measuring the elapsed time directly on the components.


\begin{example} We have implemented a simple heating control, 
where a resistor controlled by the node 1  heats up to 30 degrees 
and a fan is activated unless one of the two safety buttons controlled 
by node 0 are not pressed. We can specify the correct behavior using
the following formula: 
$$\square( (!b_0 \vee !b_1) \wedge ((t > 30) \rightarrow (fan_{on})) )$$
\end{example}

\begin{figure}[h!]
  \centering
       \includegraphics[width=0.4\textwidth]{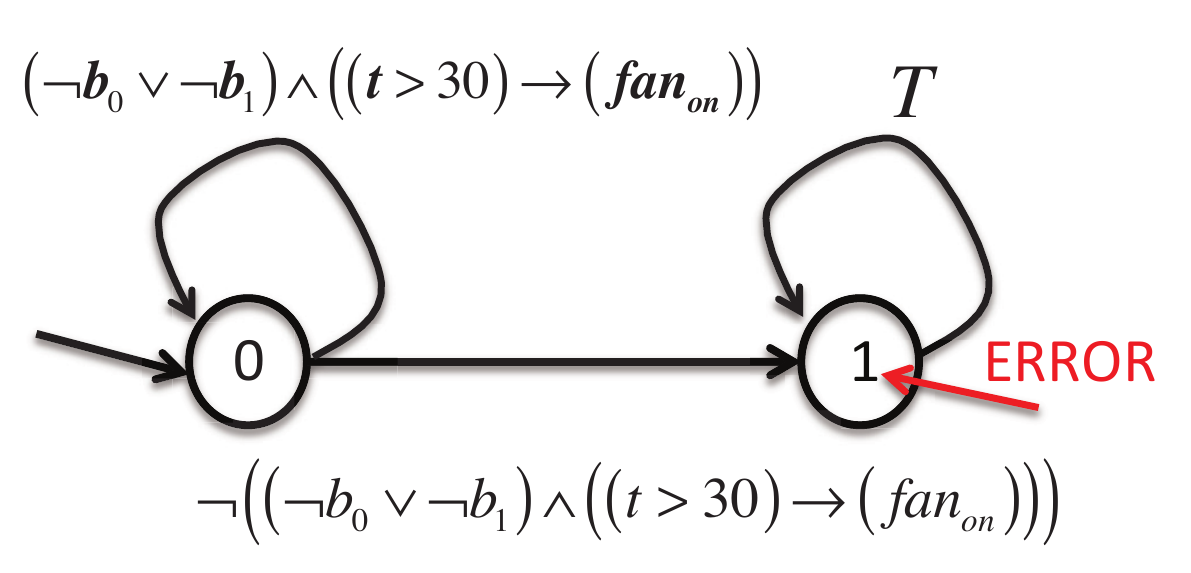}
        
   \vspace{-1mm}
  \caption{Automata-based monitor.}
  \label{fig:casestudy_monitor}
\end{figure}

\noindent The size for the automata-based monitor is only of two states as 
Figure~\ref{fig:casestudy_monitor} shows. 
The monitor based on formula progression $R(\square( (!b_0 \vee !b_1) \wedge ((t > 30) \rightarrow (fan_{on})) ), \sigma)$  will rewrite the formula as follows:

$$R((!b_0 \vee !b_1) \wedge ((t > 30) \rightarrow (fan_{on})), \sigma) \wedge \square( (!b_0 \vee !b_1) \wedge ((t > 30) \rightarrow (fan_{on})) )$$

\noindent Hence, if the term $(!b_0 \vee !b_1) \wedge ((t > 30) \rightarrow (fan_{on}))$ is true, given the events in $\sigma$, then the resulting formula is:

$$ \top \wedge \square( (!b_0 \vee !b_1) \wedge ((t > 30) \rightarrow (fan_{on})) = \square( (!b_0 \vee !b_1) \wedge ((t > 30) \rightarrow (fan_{on})) $$

\noindent otherwise, then the resulting formula is: 

$$ \bot \wedge \square( (!b_0 \vee !b_1) \wedge ((t > 30) \rightarrow (fan_{on})) = \bot $$

We have measured the $wcet_m$ time for the formula-progression monitoring of this 
example (that is also an upper bound for the automata-based monitor) counting the max 
number of CPU cycles needed with a prescaler (that divides the clock frequency) value set to 8. 
Considering that the clock speed of the micro-controller is 16 MHz, 
 we have obtained that the rewriting worst case execution time for the formula 
is $65 415 \mbox{ cycles} \times (1/ (16MHz/8)) = 0.03237 \mbox{ sec.}$

 \section{Conclusion}\label{sec:conclusion}

The synchronous decentralized monitoring of a networked embedded system 
requires some important assumptions about the synchronization mechanisms 
and the minimum sampling time  to guarantee the time consistency 
among the monitored local traces. In this work we provide a possible timed 
model in UPPAAL for  a sampling-based decentralized monitoring  and we verify 
some important properties such as the \emph{liveness}, the \emph{synchronous
sampling and the frequency}.  We then provide a case study where we 
implement this timed model in our networked 
embedded systems testbed.  Currently, we plan to extend our work in two 
directions.  First, we would like to monitor properties expressed 
in more sophisticated temporal logics dealing with dense-time such 
as Metric Interval Temporal Logic (MITL)~\cite{Alur1996}.
Secondly, since the synchronous  communication becomes very computational expensive when 
the number of components increases,  we plan to provide an asynchronous 
decentralized monitoring model, based on the Lamport's notion of global time~\cite{Lamport1978}.

 \section{Acknowledgement}
We would like to thank the students Stephan Brugger, Dominik Macher and
Daniel Schachinger that contribute in the implementation of the case study. 

\nocite{*}
\bibliographystyle{eptcs}
\bibliography{bibtex}
\end{document}